\documentclass[
    a4paper,
    notitlepage,
    pre,
    onecolumn,
    superscriptaddress,
    floatfix
]{revtex4-1}
\usepackage{amsmath,amssymb}
\usepackage{lipsum}
\usepackage{bm}
\usepackage{graphicx}
\usepackage{graphics}
\usepackage{amsmath}
\usepackage{array}
\usepackage[caption=false]{subfig}
\usepackage{xcolor}
\usepackage{subfig}
\usepackage{hyperref}
\usepackage[capitalise]{cleveref}
\usepackage{textcomp}

\usepackage{epstopdf}
\usepackage{cases}
\usepackage{amssymb}
\usepackage{multirow}
\usepackage{siunitx}
\usepackage{cases}
\usepackage{tabularx}
\usepackage[top=1.1in, bottom=1.1in, left=1.0in, right=1.0in]{geometry}
\usepackage[utf8]{inputenc}
\usepackage{tikz}
\usepackage{comment}
\usepackage{lineno,hyperref}
\modulolinenumbers[5]
\usepackage{booktabs}

\begin{document}
%

\title{Semi-implicit Lax-Wendroff kinetic scheme for hydrodynamic phonon transport}
\author{Shijie Li}
\affiliation{Department of Physics, School of Sciences, Hangzhou Dianzi University, Hangzhou 310018, China}
\author{Hong Liang}
\email{Corresponding author: lianghongstefanie@163.com}
\affiliation{Department of Physics, School of Sciences, Hangzhou Dianzi University, Hangzhou 310018, China}
\author{Songze Chen}
\affiliation{TenFong Technology Company, Nanshan Zhiyuan, No. 1001, Xueyuan Avenue, Taoyuan Street, Nanshan District, Shenzhen, China}
\author{Chuang Zhang}
\email{Corresponding author: zhangc26@zju.edu.cn}
\affiliation{College of Energy Engineering, Zhejiang University, Hangzhou 310027, China}
\date{\today}

\begin{abstract}

A semi-implicit Lax-Wendroff kinetic scheme is developed for hydrodynamic phonon transport in solid materials based on the Boltzmann transport equation under the double relaxation time approximation, in which both the normal and resistive scattering processes are accounted.
The trapezoidal and midpoint rules are adopted for the temporal integration of the scattering and migration terms under the framework of finite volume method, respectively.
Instead of direct numerical interpolation, the kinetic equation is solved again when reconstructing the interfacial flux, in order to realize the coupling of phonon migration and scattering within a numerical time step.
Specifically, the finite difference scheme is introduced and the second-order upwind or central schemes are used for the reconstruction of the interfacial distribution function and its spatial gradient.
Consequently, the cell size and time step of the present method could be larger than the phonon mean free path and relaxation time in the limit of small Knudsen numbers. 
Numerical tests demonstrate that the present method can accurately capture multi-scale thermal conduction phenomena within different normal or resistive scattering rates. 

\end{abstract}


\maketitle

\section{Introduction}

Heat conduction in solid materials is usually described by the classic Fourier law of thermal conduction~\cite{baron1878analytical}, which assumes a diffusive transport process with frequent momentum-destroying resistive (R) scattering. 
When the material dimension is reduced from 3D to 1D/2D~\cite{cepellotti_phonon_2015,lee_hydrodynamic_2015}, or the size is shortened from millimeters to nanometers~\cite{chang_breakdown_2008}, or the temperature is reduced from room temperature to near absolute zero~\cite{PhysRevLett.16.789,PhysRevLett_secondNaF}, or the heating frequency is increased from Hz to GHz~\cite{regner_broadband_2013}, the classic Fourier law of thermal conduction will fail~\cite{RevModPhysJoseph89}.
Instead of diffusive behaviors, the phonon transport may exhibit ballistic~\cite{Chuang17gray,SHEN2023_IJHMT} or hydrodynamic characteristics~\cite{beck1974,hydrodynamics_review_2022,chen_non-fourier_2021}.
The former happens when there is rare phonon scattering process, and the latter appears when the momentum-conserved phonon normal (N) scattering process is dominant and meanwhile the resistive scattering process is negligible. 

In the phonon hydrodynamics regime, the dominance of momentum-conserved N scattering engenders a collective behavior that is distinct from the stochastic random walk characteristic of diffusive phonon transport~\cite{cepellotti_phonon_2015,lee_hydrodynamic_2015,huberman_observation_2019,Huberman_2025,zhang_heat_2021}.
The thermal conduction no longer obeys the typical Fourier's law, but behaves like the viscous fluid flow, such as the phonon Poiseuille flow~\cite{nanoletterchengang_2018}, heat vortices~\cite{zhang_heat_2021,shang_heat_2020}, transient cooling~\cite{zhang_transient_2021} and second sound~\cite{huberman_observation_2019,lee_hydrodynamic_2015}.
In order to capture accurately these non-Fourier heat conduction phenomena, one popular method is the macroscopic non-Fourier equations, such as the Cattaneo or Guyer-Krumhansl equation, phase lag or kinetic-collective model~\cite{cattaneo1948sulla,tzou1995,WangMr15application,PhysRev_GK,PhysRev.148.766}.
Some high order terms are introduced in these equations to capture the time delay, nonlocal effects or nonlinear constitutive relationship between the temperature and heat flux.
However, most of them are limited in small Knudsen number, which is defined as the ratio of mean free path to system characteristic length.
Another method is the kinetic equation derived from the statistics physics~\cite{particles_statistical_physics_MIT_2007}.
Considering both computational efficiency and accuracy, the phonon Boltzmann transport equation (BTE) under the double relaxation time Callaway model~\cite{PhysRev_callaway,chuang2021graded} is a good choice to simultaneously and correctly capture the ballistic, hydrodynamics and diffusive phenomena.
However, it is difficult to use a unified numerical method to efficiently solve this model due to a wide range of Knudsen numbers of normal or resistive scattering process.

Many numerical methods have been developed to solve the BTE, such as the statistical Monte Carlo method~\cite{PhysRevB.99.085202,PhysRevLett.127.085901,SHEN2023_IJHMT} and the deterministic discrete ordinate method~\cite{Chuang17gray,wangmr17callaway}.
In the first method, many statistics particles are introduced and their evolution processes are traced one by one under the Lagrange framework.
This method can deal with complex multi-field coupling heat transfer problems, where the macroscopic variables are obtained by statistical averaging of each particle information.
However, its cell size or time step is limited by mean free path or relaxation time.
Consequently, it suffers from expensive computational cost in the diffusive or hydrodynamic regime due to the frequent scattering.
Discrete ordinate method faces a similar time-step dilemma.
This method is free of statistical noise because it directly discretizes time, position and momentum space, and usually uses direct interpolation to handle the interfacial distribution function.
However, particle migration and scattering processes are separate at a numerical time step scale, which limits its time step to the relaxation time.

To break this time-step dilemma, a semi-implicit Lax-Wendroff kinetic scheme is developed in this paper motivated by previous papers~\cite{csz_Lax_Wendroff2022,PENG202572}.
This method is based on the finite volume method framework, and the finite difference techniques are invoked to reconstruct the interfacial distribution function. 
Conventional second-order interpolation or differentiation methods are used to handle the interfacial distribution function and its spatial gradients.
Numerical results demonstrate that the present method can accurately capture the ballistic, hydrodynamics and diffusive thermal transport phenomena across a wide range of Knudsen numbers.

\section{Phonon hydrodynamics model}

To accurately predict heat conduction in three-dimensional materials, the phonon BTE under the Callaway approximation is used~\cite{PhysRev_callaway,nanoletterchengang_2018,zhang_transient_2021,chuang2021graded,WangMr15application,zhang_heat_2021,shang_heat_2020}, 
\begin{align}
\frac{\partial f}{\partial t}+\bm{v_g} \cdot \nabla_{\bm{x}} f =\frac{1}{\tau_R } \left(  {f}^{eq}_{R} - f \right)+\frac{1}{\tau_N } \left(  {f}^{eq}_{N} - f \right)
\end{align}
where $f=f(\bm{x},\bm{v_g},t)$ is the phonon distribution function of energy density, depending on spatial position $\bm{x}$, time $t$ and group velocity $ \bm{v_g} = \left|\bm{ v_g } \right| \bm{s} $, $\bm s$ is the unit directional vector and assumed to be isotropic for three-dimensional materials.
$\tau_R$ and $\tau_N$ are the relaxation times of the R and N scattering processes, respectively. 
${f}^{eq}_{R}$ and ${f}^{eq}_{N}$ represent the equilibrium distribution functions of the R and N scattering processes, respectively,
\begin{align}
{f}^{eq}_{R}=\frac{C(T-T_0)}{4 \pi},
\label{eq:feq_R}
\end{align}
\begin{align}
{f}^{eq}_{N}=\frac{C(T-T_0)}{4 \pi}+\frac{CT}{4 \pi} \frac{\bm{s}\cdot\bm{u}}{\left|\bm{ v_g } \right|},
\label{eq:feq_N}
\end{align}
where $C$ is the specific heat, $\bm{u}$ is the macroscopic drift velocity, $T$ is the macroscopic temperature and $T_0$ is the reference temperature. 
This paper focuses more on innovation in numerical formatting techniques, so it only considers gray models~\cite{zhang_transient_2021,JAPbingyangcao2022,nie2020thermal,WangMr15application,zhang_heat_2021,shang_heat_2020} for the time being.
Two Knudsen numbers ${Kn}_R$ and ${Kn}_N$ are introduced and defined as the ratio between the phonon mean free path ($\lambda_R=|\bm{v}_g| \tau_R$,$\lambda_N=|\bm{v}_g| \tau_N$) to the characteristic length respectively, i.e., $\text{Kn}_R=\lambda_R / L$, $\text{Kn}_N=\lambda_N / L$. 
Both N and R scattering processes satisfy energy conservation,
\begin{align}
\int \frac{{f}^{eq}_{R}-f}{\tau_R } d\Omega= \int \frac{{f}^{eq}_{N}-f}{\tau_N } d\Omega=0.
\end{align}
Besides, momentum is conserved during N-scattering,
\begin{align}
\int\bm{s} \frac{{f}^{eq}_{N}-f}{\tau_N } d\Omega=0, \label{eq:Nconservation}
\end{align}
where $d\Omega$ represents the integration over the entire solid angle space.
Macroscopic quantities, namely the local energy density $U$, heat flux $\bm{q}$ and temperature $T$, are obtained by taking the moments of the distribution function,
\begin{align}
U&= \int f d\Omega , \label{eq:energy} \\
\bm{q}&= \int \bm{v_g} f d\Omega ,  \label{eq:heatflux}  \\
T &=T_0+ \frac{1 }{C}  \int f d\Omega=T_0+ \frac{U}{C}. \label{eq:temperature} 
\end{align}
Actually the implicit relationship between the macroscopic heat flux and drift velocity can be obtained by combining the above formulas~(\ref{eq:heatflux},\ref{eq:Nconservation},\ref{eq:feq_N}), i.e.,
\begin{align}
\bm{q} = \frac{1}{3} C T \bm{u}.
\end{align}

\section{Semi-implicit Lax-Wendroff kinetic scheme}

A semi-implicit Lax–Wendroff kinetic scheme is introduced to numerically solve the phonon hydrodynamics model. 
To balance clarity in presentation with generality of the problem, a quasi-2D simulation is adopted as a concrete example for subsequent elaboration,
\begin{align}
\frac{\partial f}{\partial t}+u\frac{\partial f}{\partial x}+v\frac{\partial f}{\partial y} =\frac{1}{\tau_R } \left(  {f}^{eq}_{R} - f \right)+\frac{1}{\tau_N } \left(  {f}^{eq}_{N} - f \right)
\end{align}
where $u=|v_g| \cos \theta$ and $v=|v_g| \sin \theta \cos \varphi$ are the $x-$component and $y-$component of the group velocity, respectively.
A uniform Cartesian grid is used to discretize the spatial domain and the temporal space is also discretized, where $(\Delta x,~M,~i)$ and $(\Delta y,~N,~j)$ are the (cell size, total cell number, index of cell center) in the $x-$ and $y-$ direction, respectively.
The finite volume method is used to solve the transient phonon BTE at the cell center $(x_i,y_j)$ and a temporal integral from time $t^n$ to $t^{n+1}=t^{n}+\Delta t$ is performed,
\begin{align}
\int_{{t}^{n}}^{{t}^{n+1}}\frac{\partial {f}_{i,j,k}}{\partial t}dt+\int_{{t}^{n}}^{{t}^{n+1}}\left(u_k \frac{\partial {f}_{i,j,k}}{\partial x}+v_k \frac{\partial {f}_{i,j,k}}{\partial y}\right)dt=\int_{{t}^{n}}^{{t}^{n+1}}\left(\frac{{f}_{R,i,j,k}^{eq}-{f}_{i,j,k}}{\tau_R }+\frac{{f}_{N,i,j,k}^{eq}-{f}_{i,j,k}}{\tau_N }\right)dt,
\end{align}
where ${f}_{i,j,k}^{n}$=$f({x}_{i},{y}_{j},u_{k},v_{k},{t}^{n})$ and the solid angle space is also discretized with index $k$.

The midpoint rule is applied for the temporal integration of the phonon migration term and the trapezoidal rule is employed for the temporal integration of the scattering term, 
\begin{align}
\frac{{f}_{i,j,k}^{n+1}-{f}_{i,j,k}^{n}}{\Delta t}+u_k \frac{\partial {f}_{i,j,k}^{n+1/2}}{\partial x}+v_k \frac{\partial {f}_{i,j,k}^{n+1/2}}{\partial y}&=\frac{1}{2}\left(\frac{{f}_{R,i,j,k}^{eq,n}-{f}_{i,j,k}^{n}}{\tau_R }+\frac{{f}_{R,i,j,k}^{eq,n+1}-{f}_{i,j,k}^{n+1}}{\tau_R }\right)\notag\\
&+\frac{1}{2}\left(\frac{{f}_{N,i,j,k}^{eq,n}-{f}_{i,j,k}^{n}}{\tau_N }+\frac{{f}_{N,i,j,k}^{eq,n+1}-{f}_{i,j,k}^{n+1}}{\tau_N }\right).
\label{eq:DBTEcenter}
\end{align}
The spatial divergence of the distribution function is calculated as follows,
\begin{align}
\frac{\partial {f}_{i,j,k}^{n+1/2}}{\partial x} &=\frac{{f}_{i+1/2,j,k}^{n+1/2}-{f}_{i-1/2,j,k}^{n+1/2}}{\Delta x} , \label{eq:divergencex}  \\
\frac{\partial {f}_{i,j,k}^{n+1/2}}{\partial y} &=\frac{{f}_{i,j+1/2,k}^{n+1/2}-{f}_{i,j-1/2,k}^{n+1/2}}{\Delta y}, \label{eq:divergencey}  
\end{align}
where $(i \pm 1/2,j)$ or $(i,j \pm 1/2)$ represents the indexes of cell interfaces connected to cell center $(i,j)$ in the $x-$ and $y-$ direction, respectively.

Combining above three Eqs.~(\ref{eq:DBTEcenter},\ref{eq:divergencex},\ref{eq:divergencey}) , the distribution function at the cell center at the next time step can be obtained,
\begin{align}
{f}_{i,j,k}^{n+1} &=\frac{\tau_R\tau_N-h(\tau_R+\tau_N) }{\tau_R\tau_N +h(\tau_R+\tau_N)}{f}_{i,j,k}^{n}+\frac{h}{\tau_R\tau_N +h(\tau_R+\tau_N)}\left[\tau_N({f}_{R,i,j,k}^{eq,n+1}+{f}_{R,i,j,k}^{eq,n})+\tau_R({f}_{N,i,j,k}^{eq,n+1}+{f}_{N,i,j,k}^{eq,n})\right]\notag\\
&-\frac{2h\tau_R\tau_N }{\tau_R\tau_N+h(\tau_R+\tau_N)}\left(u_k\frac{{f}_{i+1/2,j,k}^{n+1/2}-{f}_{i-1/2,j,k}^{n+1/2}}{\Delta x}+v_k\frac{{f}_{i,j+1/2,k}^{n+1/2}-{f}_{i,j-1/2,k}^{n+1/2}}{\Delta y}\right) ,
\label{eq:function_center}
\end{align}
where $h=\Delta t/2$.
Taking the moment of Eq.~\eqref{eq:DBTEcenter} leads to the macroscopic governing equation for the energy density $U$ and heat flux $q_x,q_y$ at the cell center,
\begin{align}
{U}_{i,j,k}^{n+1}&={U}_{i,j,k}^{n}-\Delta t \sum_{k} \left(u_k \frac{{f}_{i+1/2,j,k}^{n+1/2}-{f}_{i-1/2,j,k}^{n+1/2}}{\Delta x}+v_k \frac{{f}_{i,j+1/2,k}^{n+1/2}-{f}_{i,j-1/2,k}^{n+1/2}}{\Delta y}\right) {\phi }_{k}, \label{eq：W_next_center}  \\
{q}_{x,i,j,k}^{n+1}&=\frac{\tau_R-h}{\tau_R+h}{q}_{x,i,j,k}^{n}-\frac{\tau_R\Delta t}{\tau_R+h} \sum_{k} u_k\left(u_k \frac{{f}_{i+1/2,j,k}^{n+1/2}-{f}_{i-1/2,j,k}^{n+1/2}}{\Delta x}+v_k \frac{{f}_{i,j+1/2,k}^{n+1/2}-{f}_{i,j-1/2,k}^{n+1/2}}{\Delta y}\right) {\phi }_{k},\label{eq：qx_next_center}\\
{q}_{y,i,j,k}^{n+1}&=\frac{\tau_R-h}{\tau_R+h}{q}_{y,i,j,k}^{n}-\frac{\tau_R\Delta t}{\tau_R+h} \sum_{k} v_k\left(u_k \frac{{f}_{i+1/2,j,k}^{n+1/2}-{f}_{i-1/2,j,k}^{n+1/2}}{\Delta x}+v_k \frac{{f}_{i,j+1/2,k}^{n+1/2}-{f}_{i,j-1/2,k}^{n+1/2}}{\Delta y}\right) {\phi }_{k}.
\label{eq：qy_next_center}
\end{align}
where the integral is evaluated via numerical quadrature $\sum_{k}$ over the discretized solid angle space with ${\phi }_{k}$ being the associated weight.

The key to solve above Eqs.~(\ref{eq:function_center},\ref{eq：W_next_center},\ref{eq：qx_next_center},\ref{eq：qy_next_center}) is to obtain the phonon distribution function at the cell interface at the half-time step $t^{n+1/2}$. 
At the cell interfaces $(x_{i+1/2}, y_j)$ and $(x_i, y_{j+1/2})$, the phonon BTE is integrated over the time interval from $t^n$ to the half-time step $t^{n+1/2} = t^n + h$,
\begin{align}
&\int_{{t}^{n}}^{{t}^{n+1/2}}\frac{\partial {f}_{i+1/2,j,k}}{\partial t}dt+\int_{{t}^{n}}^{{t}^{n+1/2}}\left(u_k \frac{\partial {f}_{i+1/2,j,k}}{\partial x}+v_k \frac{\partial {f}_{i+1/2,j,k}}{\partial y}\right)dt \notag\\&=\int_{{t}^{n}}^{{t}^{n+1/2}}\left(\frac{{f}_{R,i+1/2,j,k}^{eq}-{f}_{i+1/2,j,k}}{\tau_R }+\frac{{f}_{N,i+1/2,j,k}^{eq}-{f}_{i+1/2,j,k}}{\tau_N }\right)dt , \\
&\int_{{t}^{n}}^{{t}^{n+1/2}}\frac{\partial {f}_{i,j+1/2,k}}{\partial t}dt+\int_{{t}^{n}}^{{t}^{n+1/2}}\left(u_k\frac{\partial {f}_{i,j+1/2,k}}{\partial x}+v_k\frac{\partial {f}_{i,j+1/2,k}}{\partial y}\right)dt \notag\\&=\int_{{t}^{n}}^{{t}^{n+1/2}}\left(\frac{{f}_{R,i,j+1/2,k}^{eq}-{f}_{i,j+1/2,k}}{\tau_R }+\frac{{f}_{N,i,j+1/2,k}^{eq}-{f}_{i,j+1/2,k}}{\tau_N }\right)dt.
\end{align}
Distinct from the characteristic line solutions utilized in discrete unified gas kinetic scheme~\cite{GuoZl16DUGKS,luo2019,eDUGKS_2024}, the finite difference method is used for the reconstruction of the interfacial distribution function in this paper~\cite{csz_Lax_Wendroff2022,PENG202572}.
Specifically, the forward Euler method is applied to the migration term, while the backward Euler method is adopted for the scattering term,
\begin{align}
&\frac{{f}_{i+1/2,j,k}^{n+1/2}-{f}_{i+1/2,j,k}^{n}}{h}+u_k\frac{\partial {f}_{i+1/2,j,k}^{n}}{\partial x}+v_k\frac{\partial {f}_{i+1/2,j,k}^{n}}{\partial y} \notag\\&=\frac{{f}_{R,i+1/2,j,k}^{eq,n+1/2}-{f}_{i+1/2,j,k}^{n+1/2}}{ \tau_R}+\frac{{f}_{N,i+1/2,j,k}^{eq,n+1/2}-{f}_{i+1/2,j,k}^{n+1/2}}{\tau_N}, \label{eq:BTEfacex} \\
&\frac{{f}_{i,j+1/2,k}^{n+1/2}-{f}_{i,j+1/2,k}^{n}}{h}+u_k\frac{\partial {f}_{i,j+1/2,k}^{n}}{\partial x}+v_k\frac{\partial {f}_{i,j+1/2,k}^{n}}{\partial y} \notag\\&=\frac{{f}_{R,i,j+1/2,k}^{eq,n+1/2}-{f}_{i,j+1/2,k}^{n+1/2}}{\tau_R} +\frac{{f}_{N,i,j+1/2,k}^{eq,n+1/2}-{f}_{i,j+1/2,k}^{n+1/2}}{ \tau_N}. \label{eq:BTEfacey}
\end{align}
Rearranging these two Eqs.~(\ref{eq:BTEfacex},\ref{eq:BTEfacey}) yields the distribution function at the cell interface at the half-time step,
\begin{align}
{f}_{i+1/2,j,k}^{n+1/2} &=\frac{\tau_R\tau_N }{\tau_R\tau_N+h(\tau_R+\tau_N)}{f}_{i+1/2,j,k}^{n}-\frac{h\tau_R\tau_N}{\tau_R\tau_N +h(\tau_R+\tau_N)}\left(u_k \frac{\partial {f}_{i+1/2,j,k}^{n}}{\partial x}+v_k \frac{\partial {f}_{i+1/2,j,k}^{n}}{\partial y}\right)\notag\\&+\frac{h}{\tau_R\tau_N+h(\tau_R+\tau_N)}\left(\tau_N{f}_{R,i+1/2,j,k}^{eq,n+1/2}-\tau_R{f}_{N,i+1/2,j,k}^{eq,n+1/2}\right) , \label{eq:functionface_x} \\
{f}_{i,j+1/2,k}^{n+1/2} &=\frac{\tau_R\tau_N}{\tau_R\tau_N+h(\tau_R+\tau_N)}{f}_{i,j+1/2,k}^{n}-\frac{h\tau_R\tau_N}{\tau_R\tau_N+h(\tau_R+\tau_N)}\left(u_k \frac{\partial {f}_{i,j+1/2,k}^{n}}{\partial x}+v_k \frac{\partial {f}_{i,j+1/2,k}^{n}}{\partial y}\right)\notag\\&+\frac{h}{\tau_R\tau_N+h(\tau_R+\tau_N)}\left(\tau_N{f}_{R,i,j+1/2,k}^{eq,n+1/2}-\tau_R{f}_{N,i,j+1/2,k}^{eq,n+1/2}\right). \label{eq:functionface_y}
\end{align}
Furthermore, taking the moment of Eqs.~(\ref{eq:BTEfacex},\ref{eq:BTEfacey}) leads to the macroscopic evolution equations at the cell boundaries,
\begin{align}
{U}_{i+1/2,j,k}^{n+1/2} &={U}_{i+1/2,j,k}^{n} - h  \sum_{k}   \left(u_k\frac {\partial{f}_{i+1/2,j,k}^{n}}{\partial x}+v_k\frac{\partial {f}_{i+1/2,j,k}^{n}}{\partial y}\right){\phi }_{k} ,
\label{eq：W_half_face_x} \\
{U}_{i,j+1/2,k}^{n+1/2} &={U}_{i,j+1/2,k}^{n} - h \sum_{k}  \left( u_k\frac {\partial{f}_{i,j+1/2,k}^{n}}{\partial x}+v_k\frac {\partial{f}_{i,j+1/2,k}^{n}}{\partial y} \right) {\phi }_{k}. \label{eq：W_half_face_y}  \\
{q}_{x,i+1/2,j,k}^{n+1/2} &=\frac{\tau_R}{h+\tau_R}{q}_{x,i+1/2,j,k}^{n} -\frac{h\tau_R} {h+\tau_R}  \sum_{k}   \left(u_ku_k\frac {\partial{f}_{i+1/2,j,k}^{n}}{\partial x}+u_kv_k\frac{\partial {f}_{i+1/2,j,k}^{n}}{\partial y}\right){\phi }_{k} ,
\label{eq：qx_half_face_x} \\
{q}_{y,i+1/2,j,k}^{n+1/2} &=\frac{\tau_R}{h+\tau_R}{q}_{y,i+1/2,j,k}^{n} -\frac{h\tau_R} {h+\tau_R} \sum_{k}  \left(v_k u_k\frac {\partial{f}_{i+1/2,j,k}^{n}}{\partial x}+v_kv_k\frac {\partial{f}_{i+1/2,j,k}^{n}}{\partial y} \right) {\phi }_{k}.
\label{eq：qy_half_face_x}\\
{q}_{x,i,j+1/2,k}^{n+1/2} &=\frac{\tau_R}{h+\tau_R}{q}_{x,i,j+1/2,k}^{n} -\frac{h\tau_R} {h+\tau_R}  \sum_{k}   \left(u_ku_k\frac {\partial{f}_{i,j+1/2,k}^{n}}{\partial x}+u_kv_k\frac{\partial {f}_{i,j+1/2,k}^{n}}{\partial y}\right){\phi }_{k} ,
\label{eq：qx_half_face_y} \\
{q}_{y,i,j+1/2,k}^{n+1/2} &=\frac{\tau_R}{h+\tau_R}{q}_{y,i,j+1/2,k}^{n} -\frac{h\tau_R} {h+\tau_R} \sum_{k}  \left(v_k u_k\frac {\partial{f}_{i,j+1/2,k}^{n}}{\partial x}+v_kv_k\frac {\partial{f}_{i,j+1/2,k}^{n}}{\partial y} \right) {\phi }_{k}.
\label{eq：qy_half_face_y}
\end{align}
Under the framework of the Callaway model, the equilibrium distribution functions of resistive scattering at the cell interface  $f^{eq,n+1/2}_R$ is determined by the energy density $U^{n+1/2}$ based on Eqs.~(\ref{eq:feq_R},\ref{eq:temperature}), while the equilibrium distribution functions of normal scattering at the cell interface $f^{eq,n+1/2}_N$ is determined by both the temperature and heat flux based on Eq.~\eqref{eq:feq_N}. 
To obtain the distribution function and its spatial gradient at the cell interface, the same processing method as previous work~\cite{PENG202572} is adopted.

In summary, the evolution process of the present method for phonon hydrodynamics can be listed as below:
\begin{enumerate}
\item The interfacial distribution function and its associated spatial gradients at the current time step $t^n$ are calculated by a second-order interpolation method, as mentioned in the previous work~\cite{PENG202572}.
\item At the half-time step $t^{n+1/2}$, the macroscopic fields at the cell interfaces are updated based on Eqs.~(\ref{eq：W_half_face_x},\ref{eq：W_half_face_y},\ref{eq：qx_half_face_x},\ref{eq：qy_half_face_x},\ref{eq：qx_half_face_y},\ref{eq：qy_half_face_y}). Then, the equilibrium states can be determined by these macroscopic variables based on Eqs.~(\ref{eq:feq_R},\ref{eq:feq_N},\ref{eq:temperature}). 
\item The distribution function at the cell interfaces at the half-time step $t^{n+1/2}$ is updated based on Eqs.~(\ref{eq:functionface_x},\ref{eq:functionface_y}).
\item Update sequentially the macroscopic fields and equilibrium state at the next time step $t^{n+1}$ at the cell center based on Eqs.~(\ref{eq：W_next_center},\ref{eq：qx_next_center},\ref{eq：qy_next_center}) and Eqs.~(\ref{eq:feq_R},\ref{eq:feq_N},\ref{eq:temperature}), respectively.
\item The distribution function at the cell center at the next time step $t^{n+1}$ is updated based on Eq.~\eqref{eq:function_center}.
\end{enumerate}

\section{Numerical tests}

Three heat conduction tests across a wide range of Knudsen numbers are implemented.
All simulations were executed using a single core of an Intel Core i7-10750H processor (2.60 GHz).
The physical time step $\Delta t$ is
\begin{align}
\Delta t = \text{CFL} \times \frac{ \{\Delta x,\Delta y \}_{\text{min}}  }{|\bm{v_g}|},
\end{align}
where $\text{CFL} \in (0,1)$ denotes the Courant-Friedrichs-Lewy number. 
Without special statements, the non-dimensional parameters are specific heat $C$ = 1, group velocity $|\bm{v_g}|$= 1 and CFL=$0.40$. 
Detailed treatments of boundary conditions are the same as the previous paper~\cite{PENG202572}.
In quasi-1D simulations, the Gauss–Legendre quadrature is used to discrete $\cos \theta \in [-1,1]$ into $N_{\theta}$ points.
In quasi-2D or 3D simulations, the Gauss–Legendre quadrature is also used to discretize the azimuthal angle $ \varphi \in [0,\pi]$ into $N_{\varphi} /2$ points due to symmetry.

\subsection{Quasi-1D heat conduction} 

\begin{figure}[htb]
     \centering
     \subfloat[$Kn_N=10^5$]{\label{Fig1a}\includegraphics[width=0.32\textwidth]{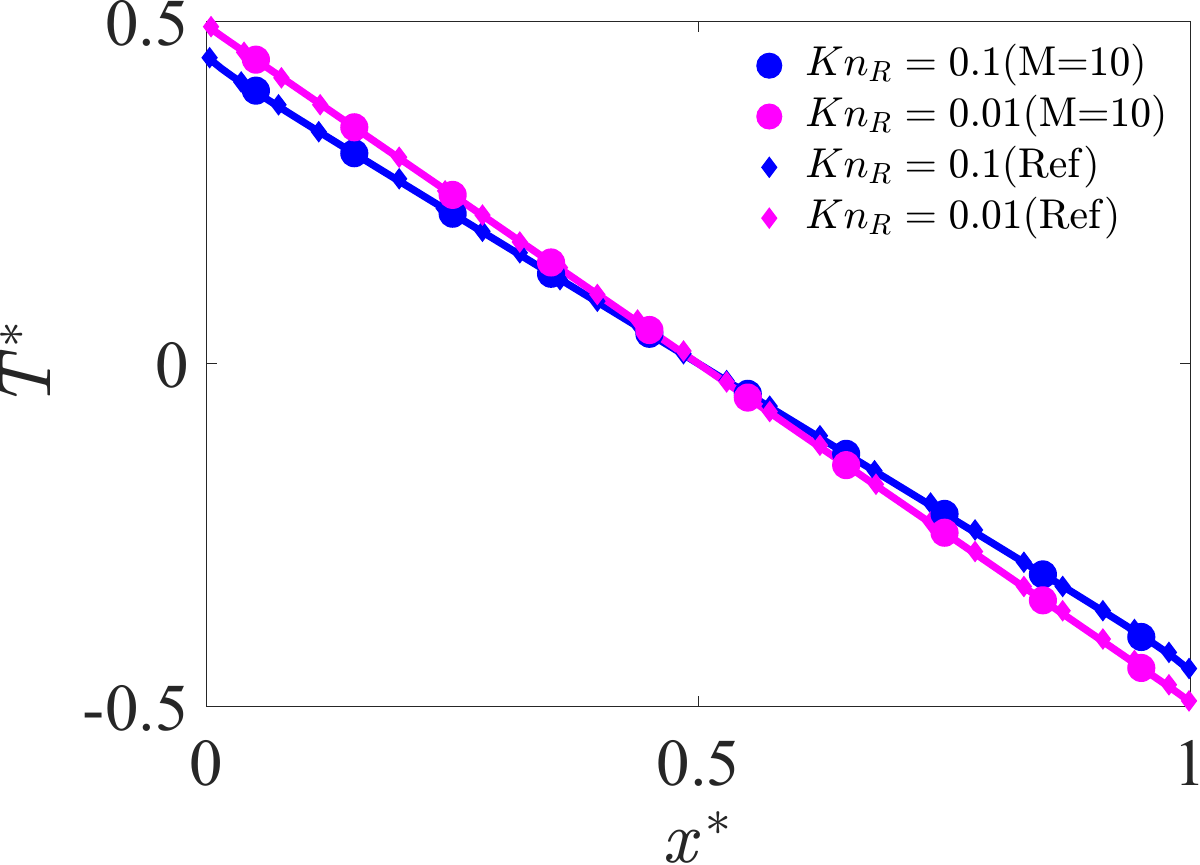}}~~
     \subfloat[$Kn_R=10^5$]{\label{Fig1b}\includegraphics[width=0.32\textwidth]{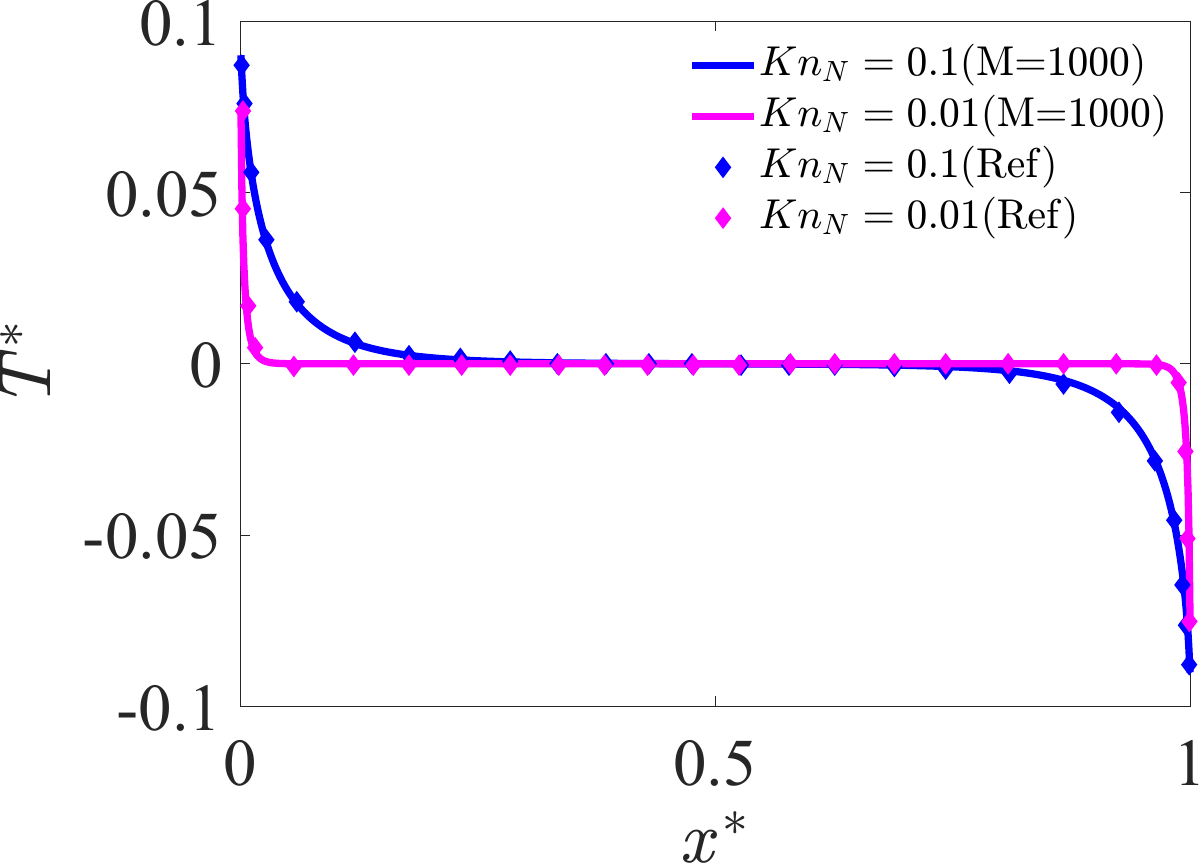}}~~ 
     \subfloat[]{\label{Fig1c}\includegraphics[width=0.32\textwidth]{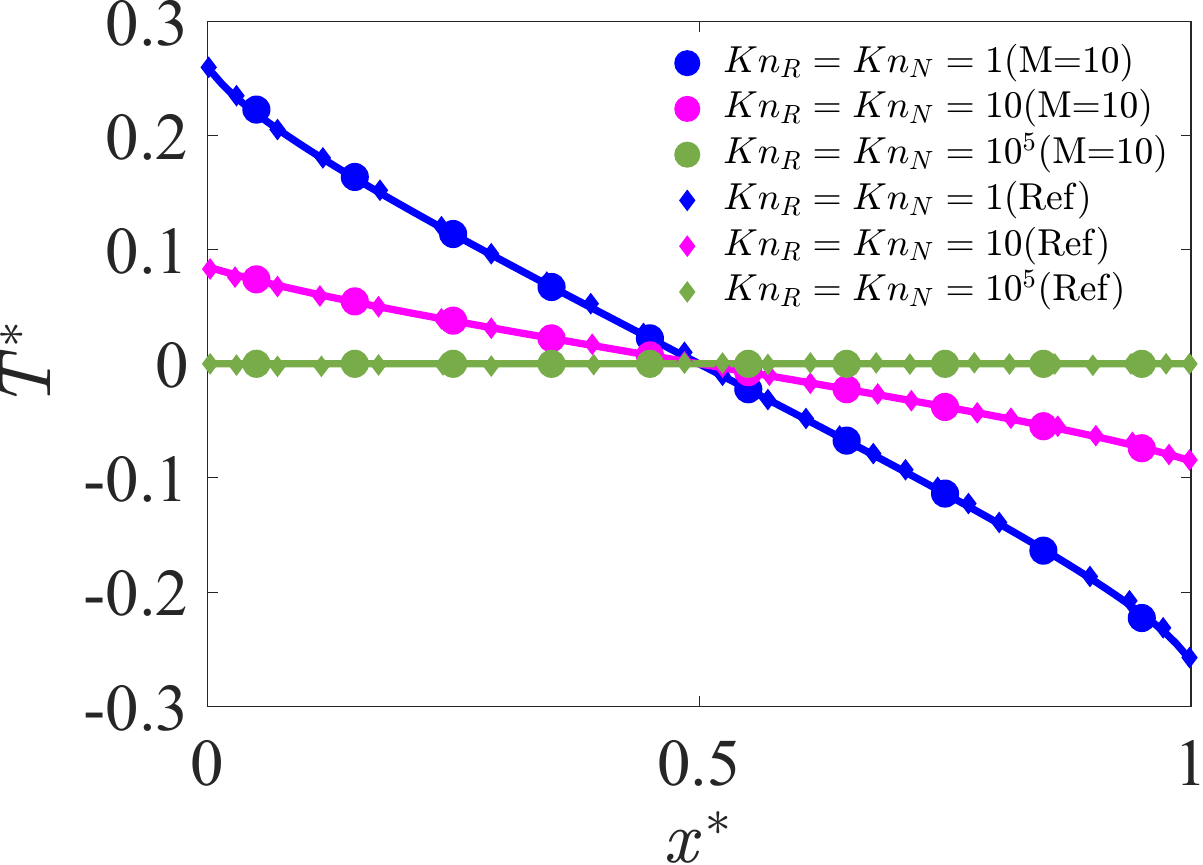}}
     \caption{Steady spatial distributions of the temperature, where ${T^*} = \left( T-T_L\right)/(T_h-T_L)$, $x^*=x/L$. The solid diamonds present the data obtained from previous references~\cite{MajumdarA93Film,Nanalytical,LIU2022111436}. The solid circles and lines present the simulated values. (a,c) $ \Delta x=0.1, \Delta t=0.04$. (b) $\Delta x=0.001, \Delta t=0.0004$.}
     \label{filmtemperature}
\end{figure}
A quasi-1D thermal transport through a thin film of unit thickness $L=1$ is investigated. 
Isothermal boundary conditions are used for two boundaries, and the left and right terminals maintained at temperatures $T_h$ and $T_c$, respectively.
Uniform meshes are used to discrete the spatial domain and $N_{\theta} =24$ is adopted for the solid angle space. 
Convergence is achieved when the criterion $\epsilon_1 < 1.0 \times 10^{-7}$ is satisfied, where
\begin{align}
\epsilon _1 = \frac{1}{M}\displaystyle\sum_{i=1}^{M}\left| \frac{{T}_{i}^{n+1}-{T}_{i}^{n}}{{T}_{i}^{n}} \right|.
\end{align}

The spatial distributions of the dimensionless temperature ${T^*} = \left( T-T_L\right)/(T_h-T_L)$ at different Knudsen numbers are plotted in~\cref{filmtemperature}, where the dimensionless coordinate is $x^*=x/L$.
As illustrated by the profiles, the dimensionless temperature distributions calculated by the present scheme demonstrate good agreement with the analytical solutions~\cite{MajumdarA93Film,Nanalytical} across the entire range from ballistic, hydrodynamic to diffusive regimes. 
The results in~\cref{Fig1a,Fig1c} show that the present scheme can obtain accurately capture macroscopic distributions even with coarse mesh ($M=10$). 
In the phonon hydrodynamic regime with $\text{Kn}=10^{-2}$ (\cref{Fig1b}), a smallest cell size ($\Delta x=10^{-3}$) has to be adopted to capture the sharp Knudsen layers and steep temperature jumps near the boundaries. 
It is important to clarify that this refinement is dictated by the physical requirement rather than by numerical stability constraints. 

\subsection{Transient thermal grating}

At the initial moment $t=0$, the temperature field within the quasi-1D domain follows a spatial cosine profile,
\begin{align}
T(x,0)=T_c+\Delta T \cos (\omega x),
\end{align}
where $T_c$ is the background temperature of the sample, $\Delta T$ is the modulation amplitude of the thermal grating satisfying $\Delta T\ll T_c$, and $\omega =2\pi /L$ is the wave number. 
The computational domain is set to be periodic and $50$ uniform meshes are used to discrete a single spatial grating period $L=1$. 

\begin{figure}[htb]
     \centering
     \subfloat[]{\label{Fig2a}\includegraphics[width=0.32\textwidth]{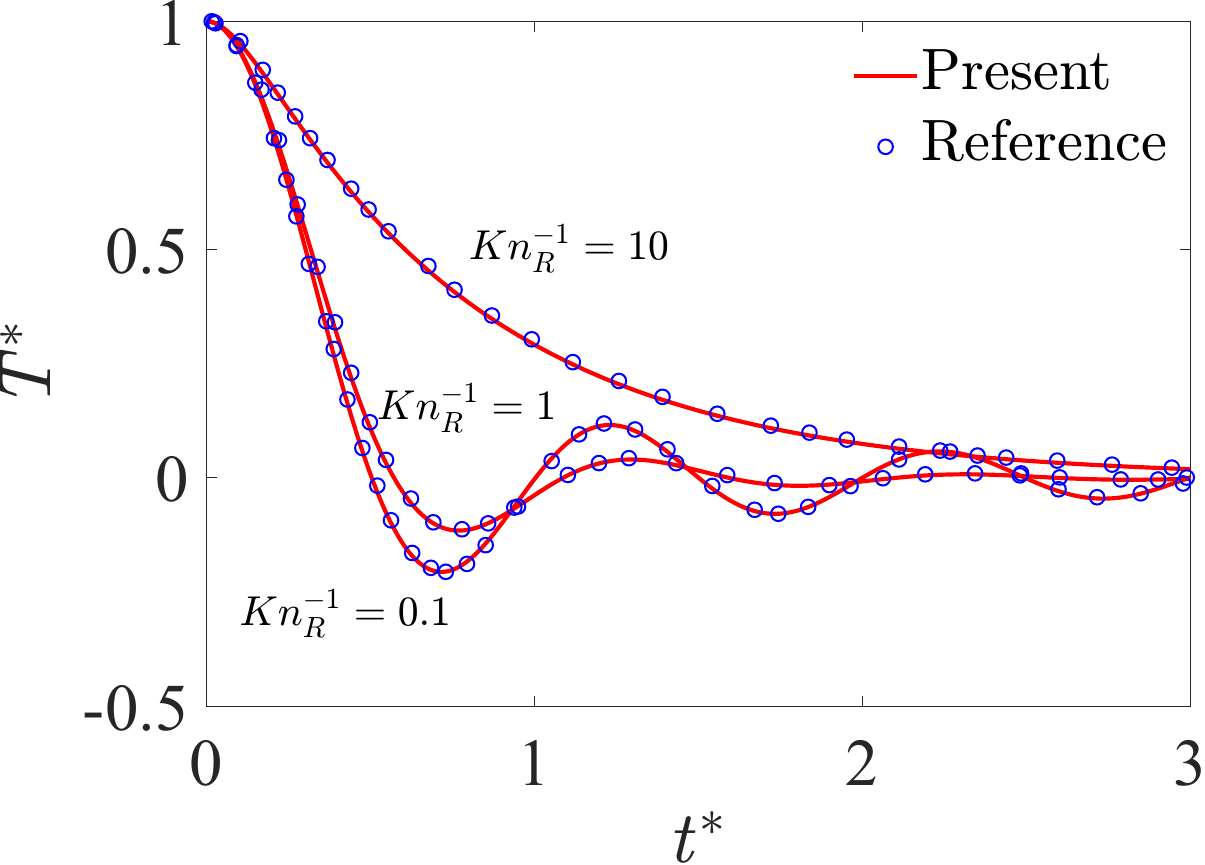}}~~
     \subfloat[]{\label{Fig2b}\includegraphics[width=0.32\textwidth]{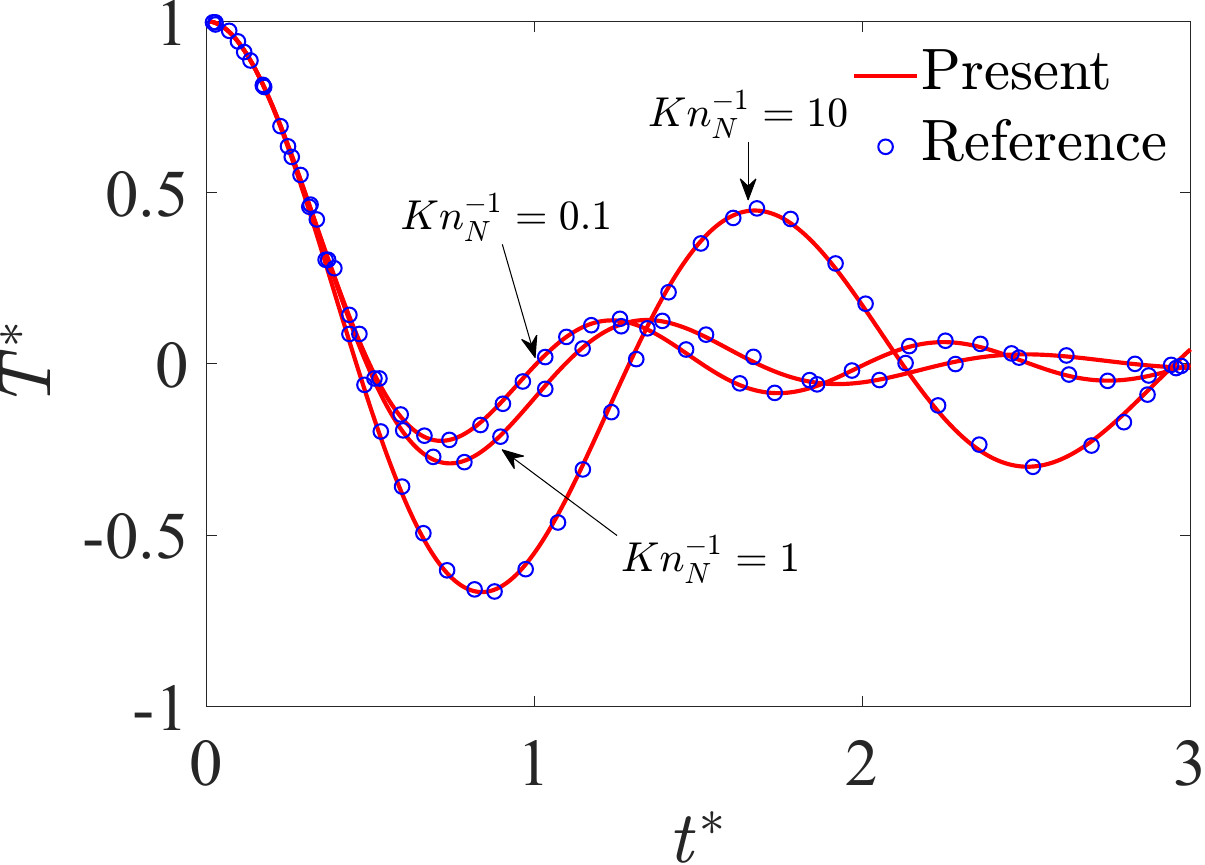}}~~ 
     \subfloat[]{\label{Fig2c}\includegraphics[width=0.32\textwidth]{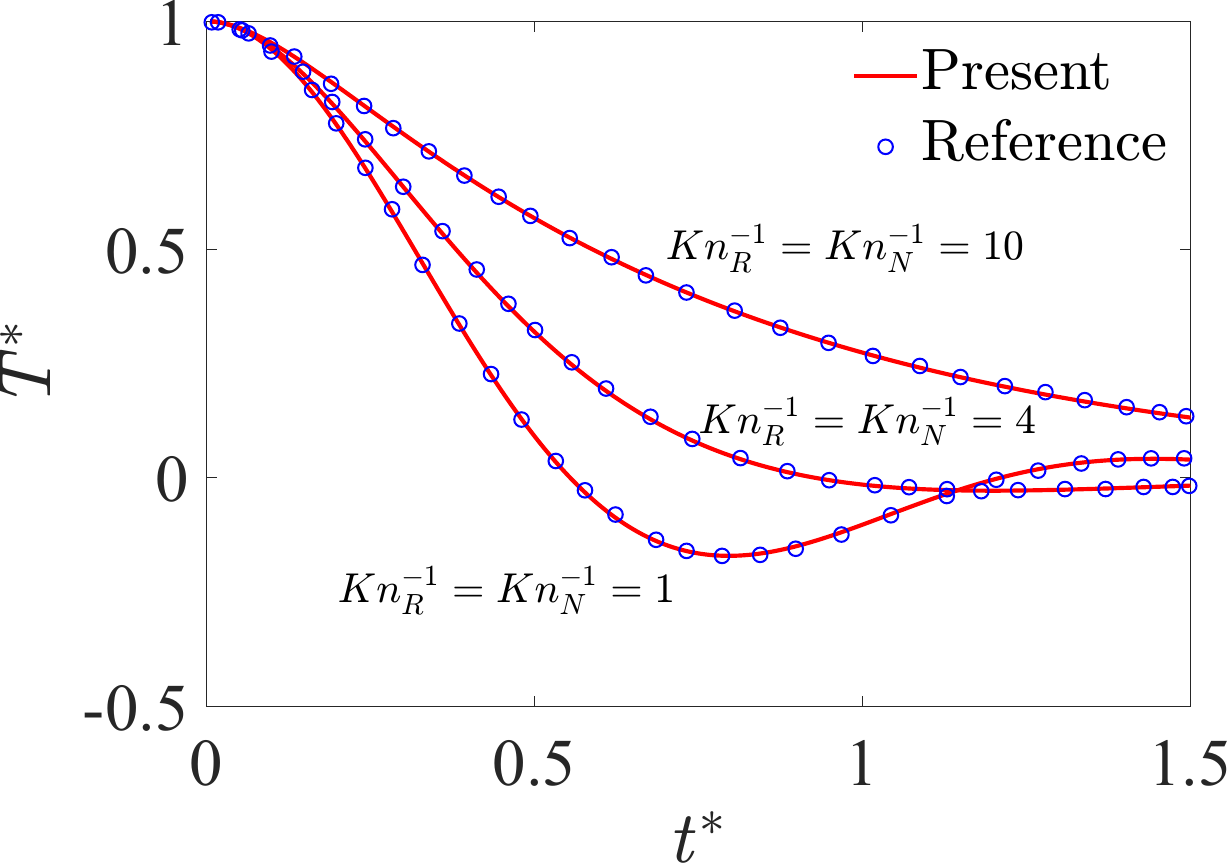}}~~
     \caption{Temporal attenuation of the temperature fluctuation amplitude across various Knudsen numbers. $\Delta x$=0.02, $\Delta t$=0.005. 'Present' is the simulated value, where $N_{\theta} \times N_{\varphi} = 48 \times 48$, `Reference' denotes the data obtained from previous papers~\cite{collins_non-diffusive_2013,heatwaves_2022chuang,PhysRevB.104.245424}. (a) Only R-scattering. (b) Only N-scattering. (c) Both R-scattering and N-scattering.}
     \label{TTG}
\end{figure}

\begin{figure}[htb]
     \centering
     \subfloat[]{\label{Fig3a}\includegraphics[width=0.4\textwidth]{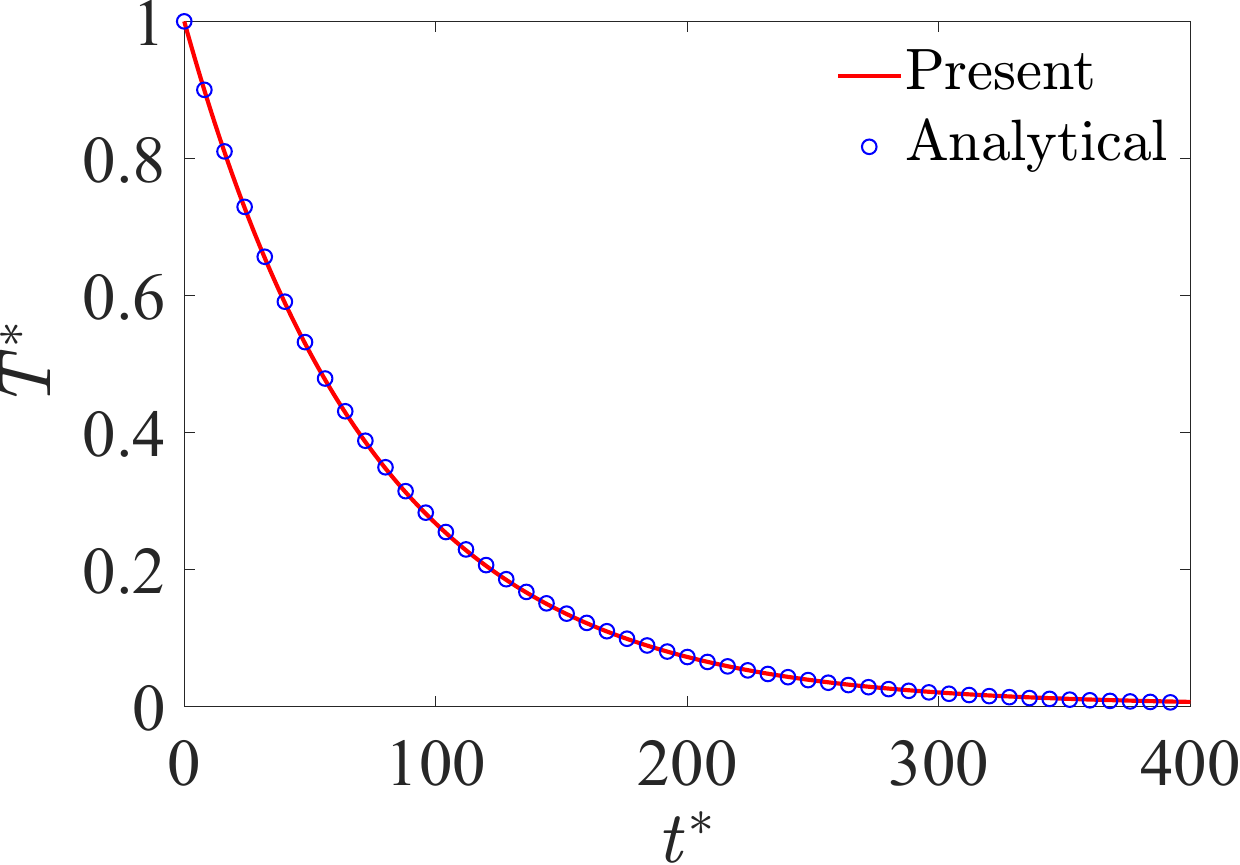}}~~
     \subfloat[]{\label{Fig3b}\includegraphics[width=0.4\textwidth]{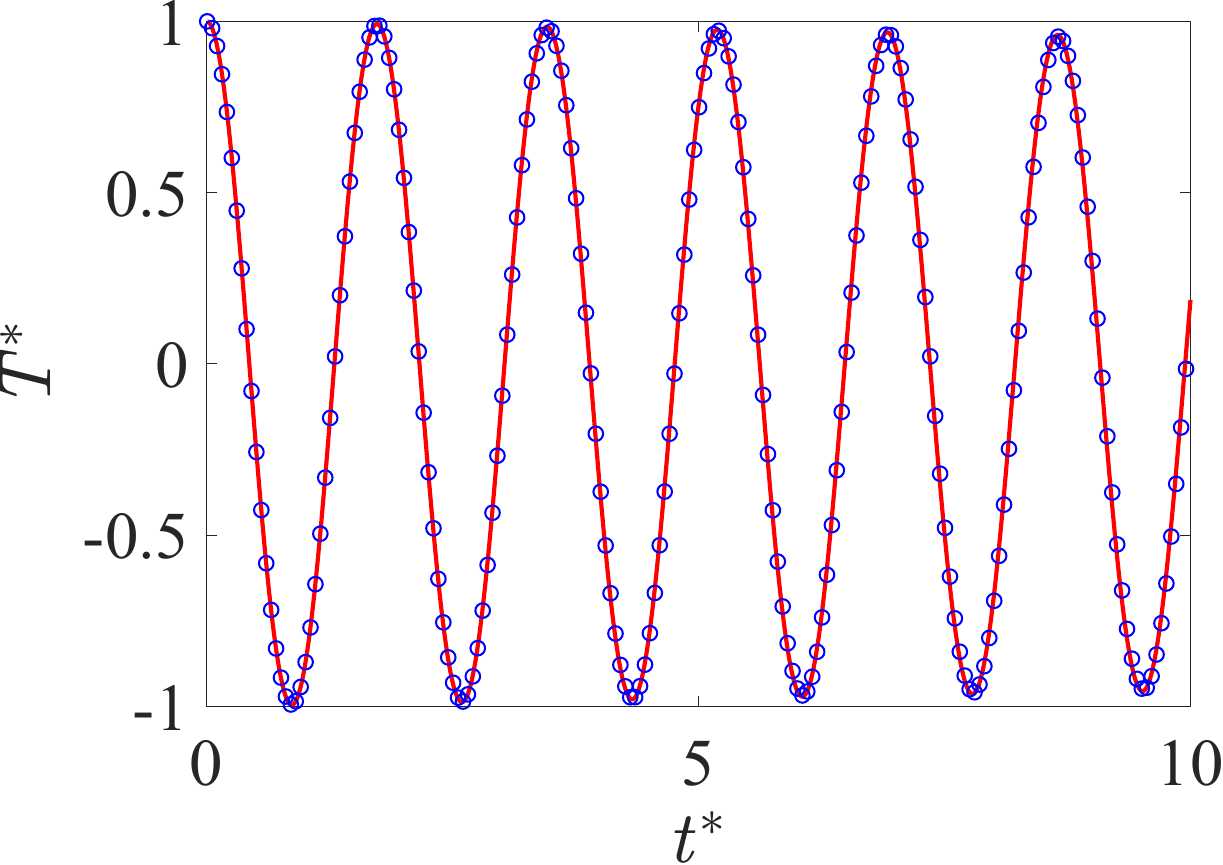}}~~\\
\caption{(a) $Kn_R=10^{-3}$, $Kn_N=10^5$, $\Delta x$/$\lambda_R$=12.5, $\Delta t$/$\tau_R$=8. Reference solutions in the diffusive regime are obtained from Ref.~\cite{collins_non-diffusive_2013} (b) $Kn_R=10^5$, $Kn_N=10^{-3}$, $\Delta x$/$\lambda_N$=10, $\Delta t$/$\tau_N$=5. Reference solutions in the hydrodynamics regime are obtained from Ref.~\cite{heatwaves_2022chuang}.}
     \label{TTG_test}
\end{figure}

Fig.~\ref{TTG} illustrates the temporal evolution of temperature at $x=0$ at various Knudsen numbers, where $T^*=(T-T_c)/\Delta T$, $t^*=v_gt/L$.
The predicted results are compared with the data obtained from previous references~\cite{collins_non-diffusive_2013,heatwaves_2022chuang}
The excellent agreement demonstrates that the present scheme could accurately describe transient heat conduction throughout the entire spectrum, from the ballistic, hydrodynamic to diffusive regime.
Furthermore, the present scheme can still accurately capture the diffusive and hydrodynamic phonon transport when the cell size and time step are much larger than the mean free path and relaxation time, as shown in Fig.~\ref{TTG_test}. 

These results presented above convincingly demonstrate that the present scheme effectively overcomes the stiffness issue of the scattering term~\cite{PhysRevE.107.025301}, liberating the time step from the constraint of relaxation times. 
Especially, the scheme accurately reproduces the wave propagation characteristics of the second sound~\cite{RevModPhysJoseph89,heatwaves_2022chuang} in the hydrodynamic regime (\cref{Fig3b}), even if the cell size and time step are significantly larger than the phonon mean free path and relaxation time. 

\subsection{Quasi-2D in-plane heat conduction} 
\begin{figure}[htb]
     \centering
     \subfloat[$Kn_R=1.0, Kn_N=10^5$]{\includegraphics[width=0.32\textwidth]{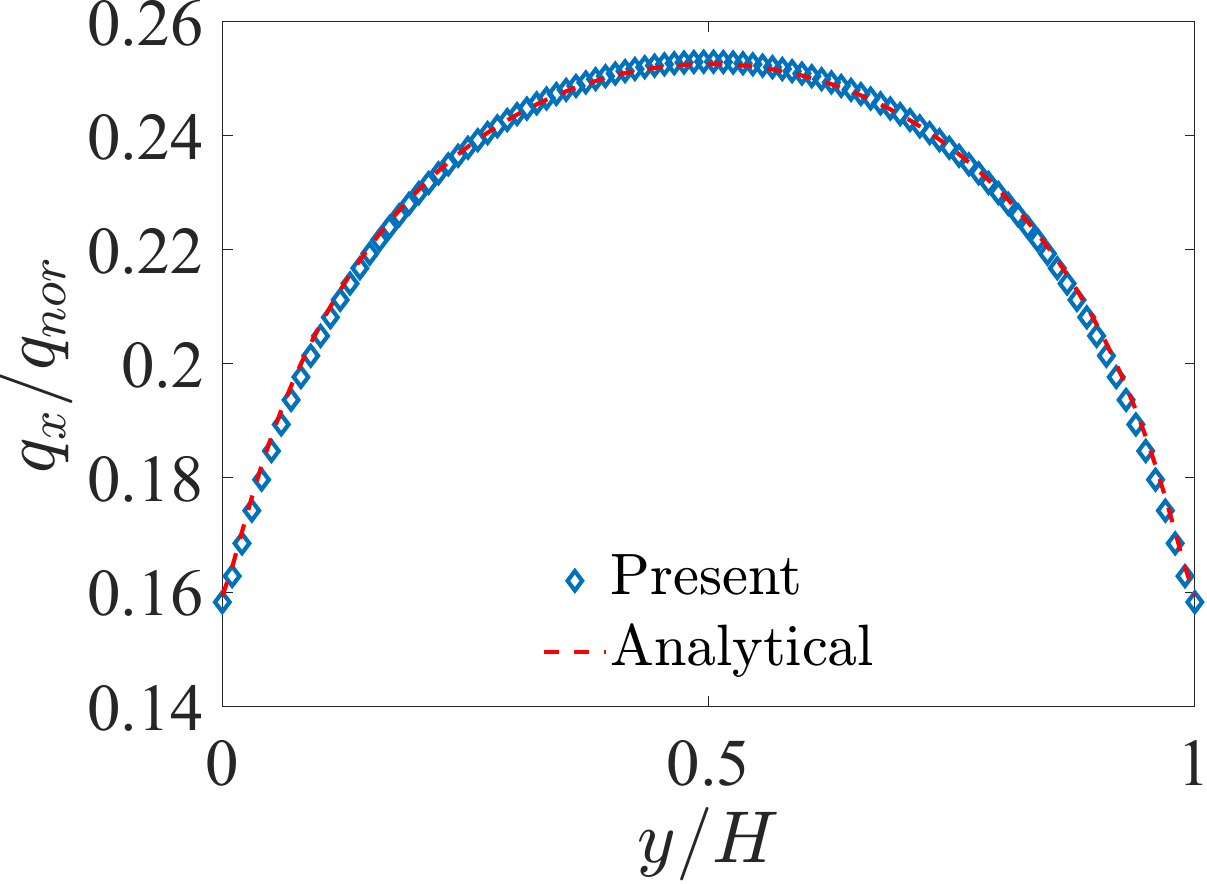}}~~
     \subfloat[$Kn_R=0.1, Kn_N=10^5$]{\includegraphics[width=0.32\textwidth]{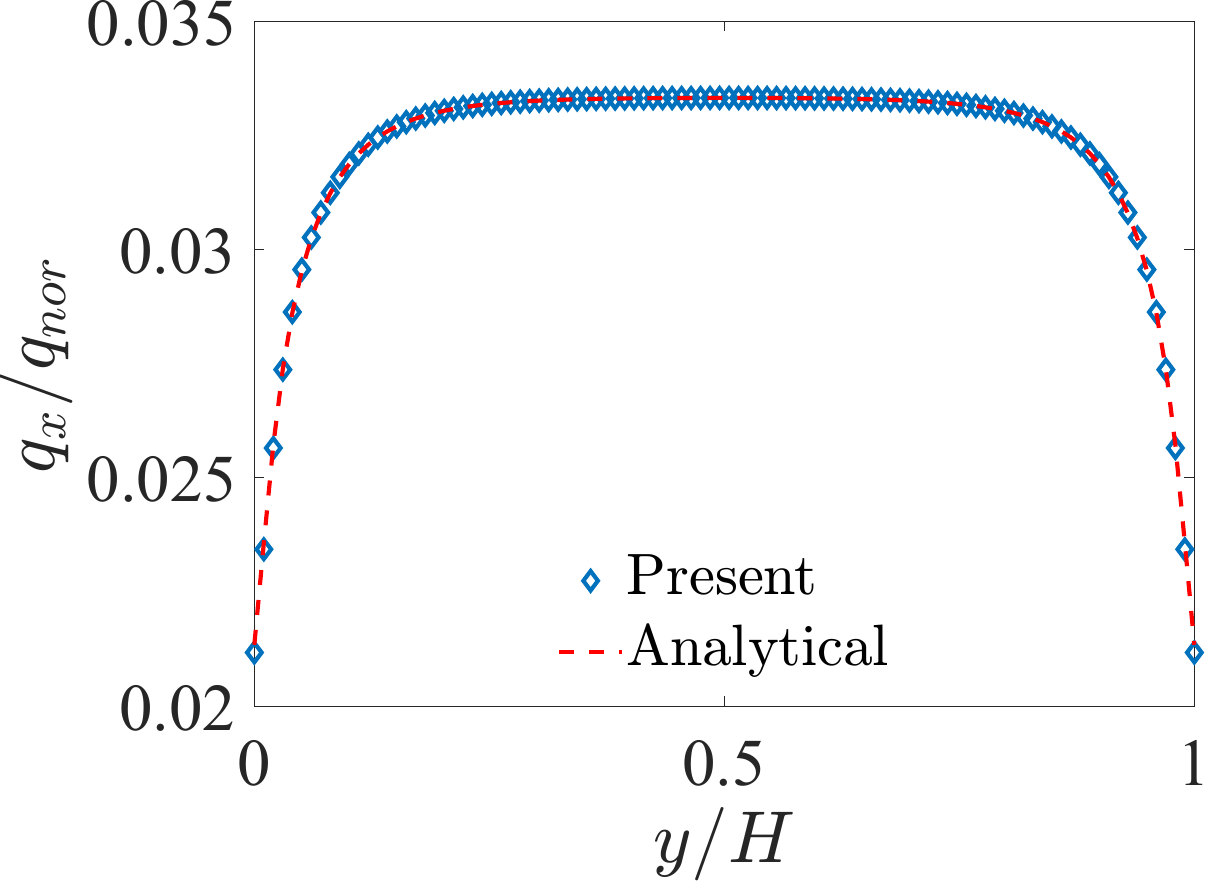}}~~
     \subfloat[$Kn_R=0.01, Kn_N=10^5$]{\includegraphics[width=0.32\textwidth]{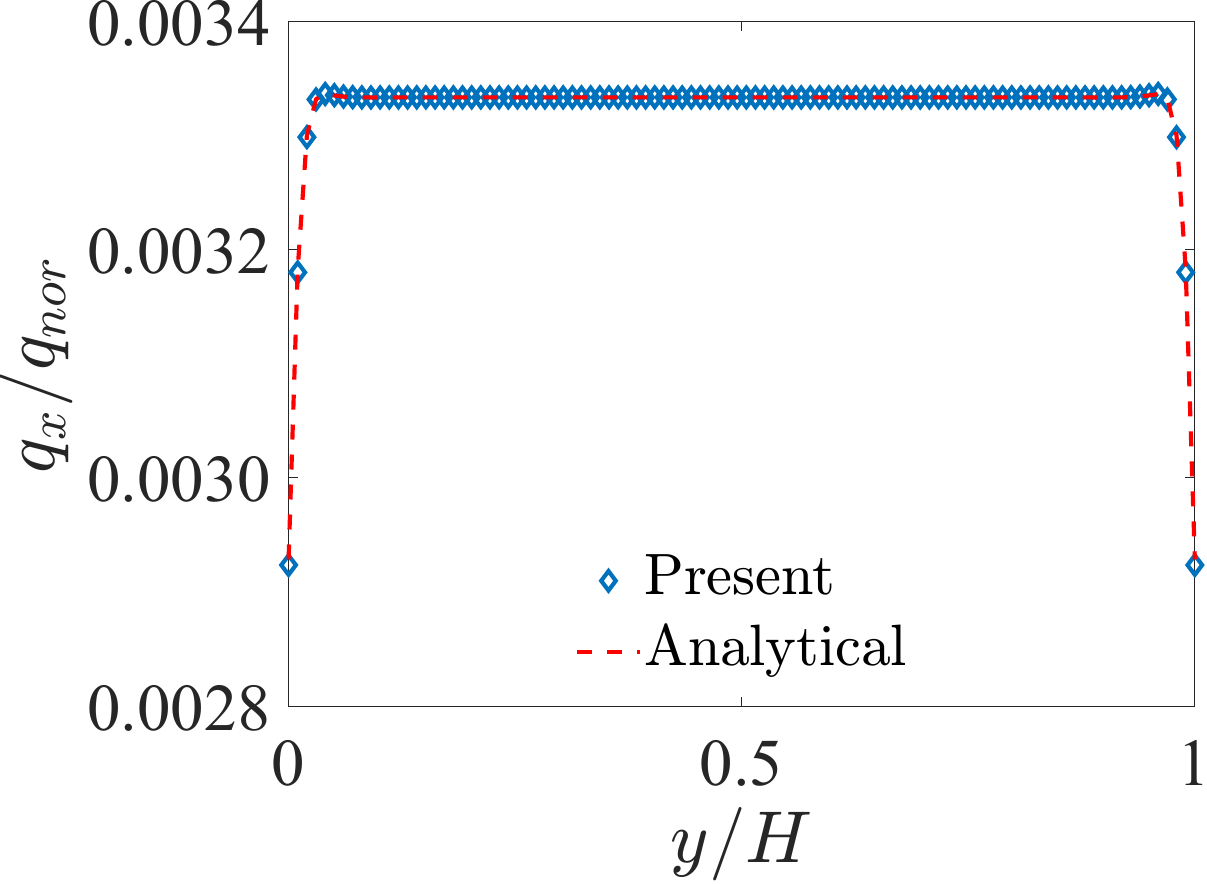}}  \\
     \subfloat[$Kn_R=10^5, Kn_N=1.0$]{\includegraphics[width=0.32\textwidth]{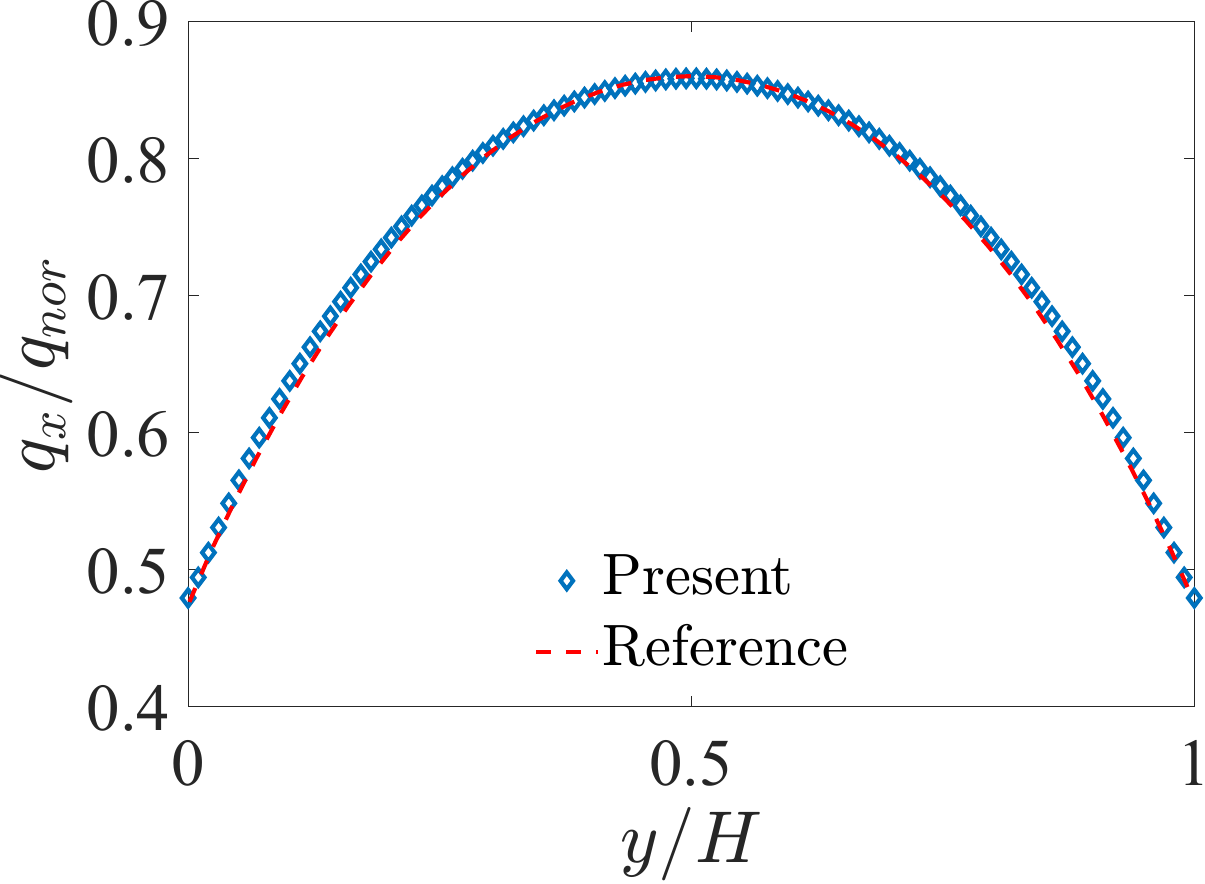}}~~
     \subfloat[$Kn_R=10^5, Kn_N=0.1$]{\includegraphics[width=0.32\textwidth]{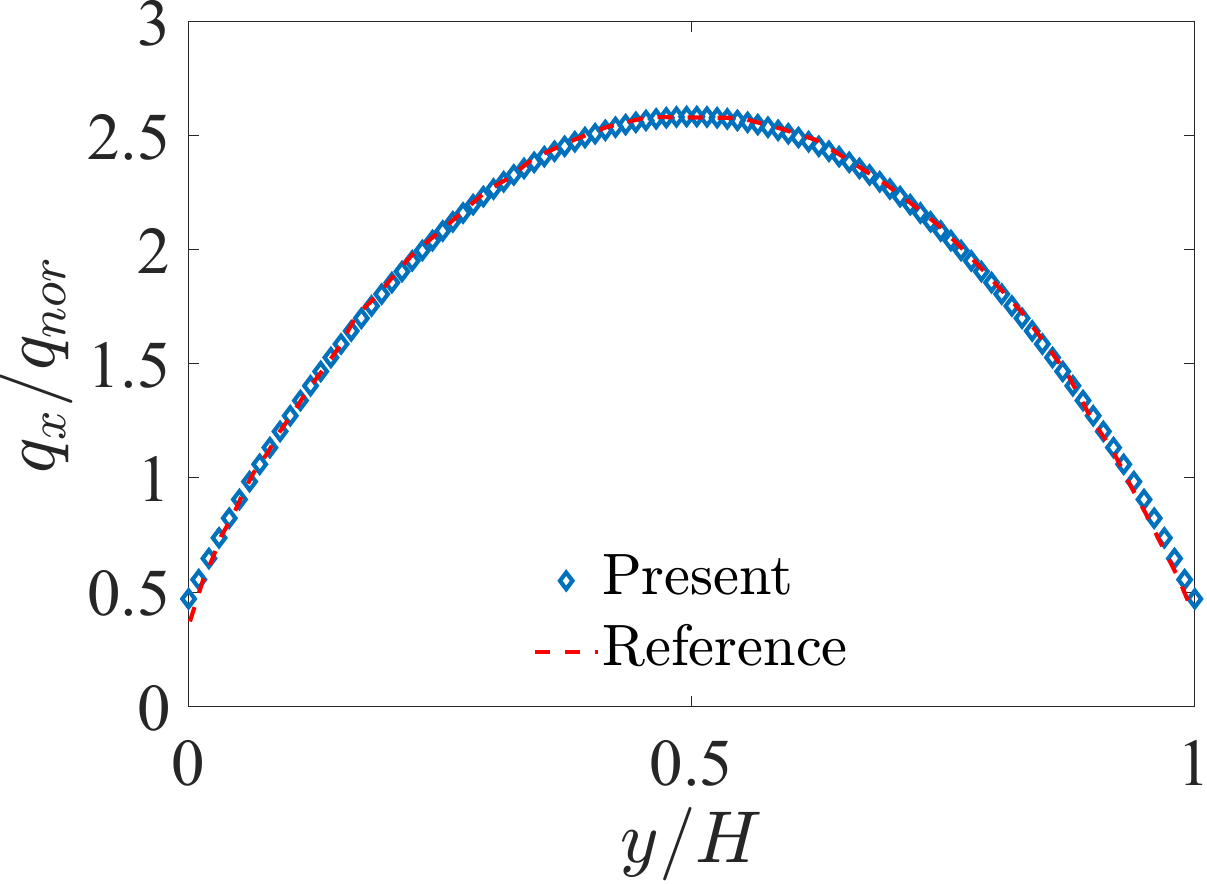}}~~
     \subfloat[$Kn_R=10^5, Kn_N=0.01$]{\includegraphics[width=0.32\textwidth]{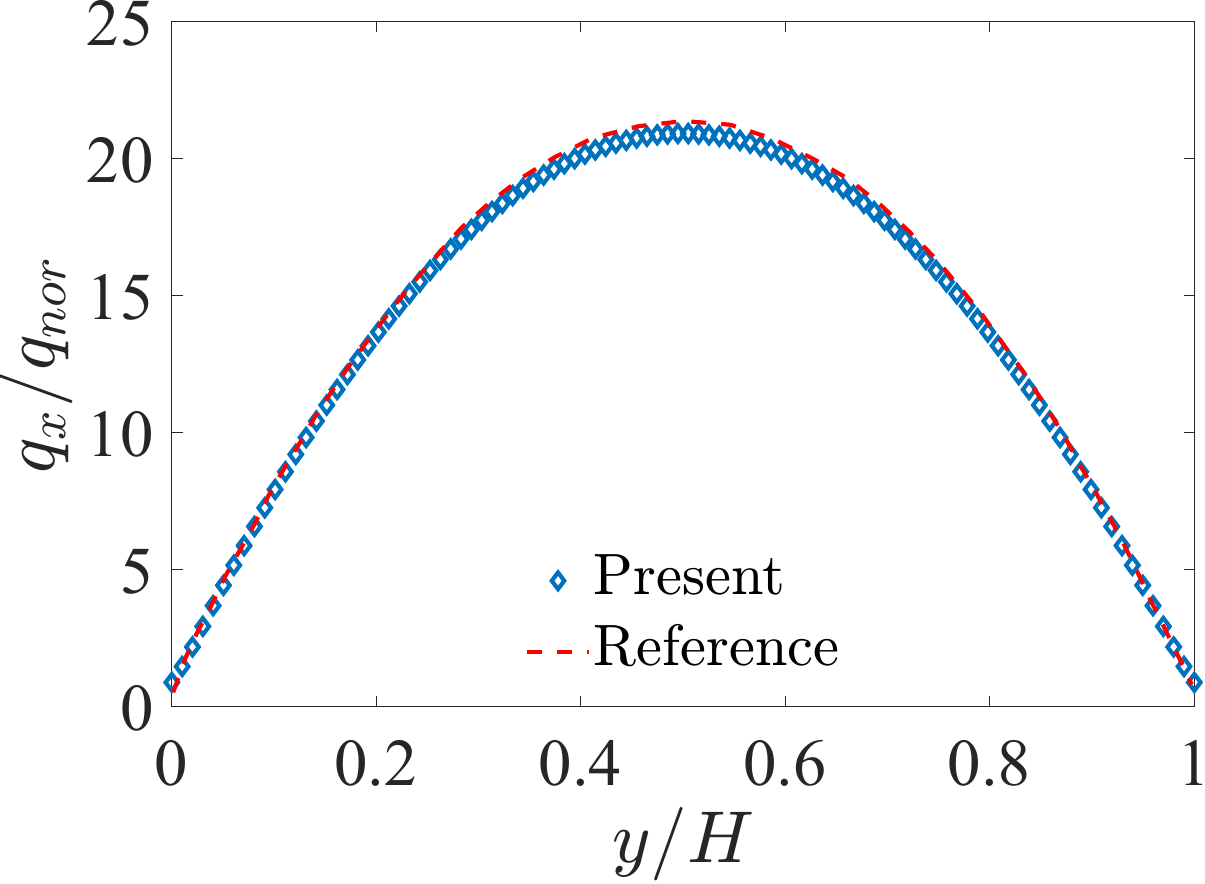}}~~
     \caption{Spatial profiles of heat flux in the diffusive regime, where $N_{\theta} \times N_{\varphi} = 24 \times 24$, $\Delta y$=0.01, $\Delta t$=0.004, `Analytical' is the Fuchs-Sondheimer analytical solutions~\cite{Fuchs_1938_analytical,Sondheimer_1952_analytical}. `Reference' is the data obtained from a previous paper~\cite{LIU2022111436}. }
     \label{h2d1}
\end{figure}

Quasi-2D in-plane heat conduction simulation is performed under a constant horizontal temperature gradient $dT/dx$.
The computational domain spans a height $H$ and a length $L$, with the system initialized at a temperature $T_c$. The temperatures in the left and right extremities are prescribed as $T_c + dT/dx \times L/2 $ and $T_c -dT/dx \times L/2$, respectively. 
The top and bottom surfaces are treated as diffusely reflecting adiabatic.
Periodic boundary conditions are applied to the left and right sides.
The computational domain is discretized into $N \times M = 100 \times 2$ uniform cells.
Convergence to steady state is achieved once the criterion $\epsilon _2<1.0\times 10^{-8}$ is satisfied,  where
\begin{align}
\epsilon _2= \frac{1}{N}\displaystyle\sum_{j=1}^{N}\left | \frac{{q}_{j}^{n+1}-{q}_{j}^{n}}{{q}_{j}^{n}}\right |,
\end{align}

The spatial profiles of heat flux are shown in~\cref{h2d1}, where $q_{nor}=-C|{\bm{v_g}}|^ 2 dT/dx$ is the reference heat flux.  
The simulation results demonstrate excellent consistency with the analytical solutions~\cite{Fuchs_1938_analytical,Sondheimer_1952_analytical} or previous data~\cite{LIU2022111436}.
It demonstrates that the present scheme can accurately describe the multi-scale heat conduction phenomena.

\section{Conclusion}

A semi-implicit Lax-Wendroff kinetic scheme is developed for solving the phonon BTE under the Callaway model. 
The most significant feature of this method is that it uses the classical finite difference method to solve the phonon BTE again and introduces second order interpolation to handle the distribution function and its spatial divergence at the cell interface.
This implement couples the phonon N-process, R-process and migration together at a single numerical time step, so that the time step could be much larger than the relaxation time when simulating diffusive or hydrodynamic heat conduction. 
Numerical results demonstrate that this method can accurately characterize steady or transient ballistic, diffusive and hydrodynamic heat conduction.

\section*{Acknowledgments}

C. Z. acknowledges the support of the National Natural Science Foundation of China (52506078) and Zhejiang Provincial Natural Science Foundation of China under Grant No.LMS26E060012.
H. L. acknowledges the support of the National Natural Science Foundation of China (12572285).
C. Z. acknowledges the members of online WeChat Group: Device Simulation Happy Exchange Group, for the communications on phonon BTE simulations.
The authors acknowledge Beijing PARATERA Tech CO., Ltd. for the HPC resources.

\bibliography{phonon}
\end{document}